\documentclass[prd,aps,twocolumn,floats,floatfix,nofootinbib] {revtex4-1}
\usepackage{booktabs}
\usepackage{graphicx}
\usepackage{dcolumn}
\usepackage{bm}
\usepackage{graphics}
\usepackage{slashed}
\usepackage{amssymb}
\usepackage{natbib}
\usepackage{amsmath}
\usepackage{url}
\usepackage[usenames,dvipsnames,svgnames,table]{xcolor}
\usepackage{color}
\usepackage{xcolor}
\usepackage{tabularx}
\usepackage{hyperref}
\usepackage{placeins}

\begin{document}

\title{Gravitational Waves from Dark Boson Star binary mergers}

\author{
Miguel Bezares$^{1},$
Carlos Palenzuela$^{1}$
}

\affiliation{${^1}$Departament  de  F\'{\i}sica $\&$ IAC3,  Universitat  de  les  Illes  Balears  and  Institut  d'Estudis
Espacials  de  Catalunya,  Palma  de  Mallorca,  Baleares  E-07122,  Spain}

\begin{abstract}
Gravitational wave astronomy might allow us to detect the coalescence of low-brightness astrophysical compact objects which are extremely difficult to be observed with current electromagnetic telescopes. Besides classical sources like black holes and neutron stars, other candidates include Exotic Compact Objects (ECOs), which could exist in theory but have never yet been observed in Nature. Among  different possibilities, here we consider Dark Stars, astrophysical compact objects made of dark matter such that only interact with other stars through gravity.  We study numerically the dynamics and the gravitational waves produced during the binary coalescence of Dark Stars composed by bosonic fields. 
These results are compared both with Post-Newtonian approximations and with previous simulations of binary boson stars, which interact both through gravity and matter.
Our analysis indicates that Dark Boson Stars belong to a new kind of compact objects, representing stars made with different species, whose merger produces a gravitational signature clearly distinguishable from other astrophysical objects like black holes, neutron stars and even boson stars.

\end{abstract}

\maketitle

\section{Introduction}\label{intro}

The recent direct detection of gravitational waves (GWs) by using LIGO and VIRGO interferometer observatories, consistent with the merger of binary black hole (BH) systems~\cite{GW150914,GW151226,GW170104,GW170608,GW170814},  has opened a new era of gravitational wave (GW) astronomy leading to unprecedented discoveries. More recently, the GW corresponding to the coalescence of a binary composed by neutron stars (NSs) has also been observed~\cite{PhysRevLett.119.161101,2017ApJ...848L..12A}. This signal was followed by a plethora of electromagnetic (EM) counterparts, including a gamma-ray burst~\cite{GWGR} and a thermal infrared/optical spectra consistent with a kilonova~\cite{Kila}, starting a fruitful era of multi-messenger astronomy. These EM and GW observations, which will be soon routinely detected, are inevitably leading to breakthroughs in our understanding of some of the most exciting objects and phenomena in the Universe, as well as providing clues to fundamental physics, such as the properties of matter at nuclear densities and stringent tests of general relativity (GR) (see for instance~\cite{Yunes:2016jcc,2016ApJ...818L..22A}). 

Nowadays, the only known astrophysical compact objects able to source strong gravitational waves are BHs and NSs~\cite{roadmap}. Nevertheless, there might be other non-standard low-brightness stars, known generically as exotic compact objects (ECOs) (see~\cite{Cardoso:2017cqb} for a review), too dim to be observed by current EM telescopes. However, if they are massive and compact  enough, it might be possible to detect them trough the GWs radiated during their coalescence.

Among the zoo of  ECOs, one of the most plausible are boson stars (BSs). BSs are solutions of Einstein equations coupled to a complex scalar field that represent a self-gravitating Bose-Einstein condensate. This family of solutions yields to useful models of dark matter, BH mimickers and simple generic compact objects (see~\cite{schumie,liebpa} for a review). Interestingly, some families of BSs can reach a compactness $C\equiv M/R$ comparable to NSs.
There are two theoretical arguments that supports the possibility of self-gravitating objects made by bosonic particles in the Universe. First, the discovery of Higgs boson~\cite{Chatrchyan:2012xdj,Aad:2012tfa} confirmed the existence of scalar fields in nature.
Second, the existence of a formation mechanism, dubbed as gravitational cooling~\cite{seidel}, to produce BSs from a generic scalar field configuration. Formation of BS has been considered also in the Newtonian limit~\cite{guzman} and much more recently for vector fields~\cite{giov}. 

The merger of binary BSs has already been studied in several works: head-on and orbital collisions of mini-BSs~\cite{pale1,pale2}, head-on collision of oscillatons (a solution of BS with real scalar field)~\cite{brito,helfer}, head-on collision of Proca stars (a solution modeled by a massive complex vector field)~\cite{sanchis} and orbital collision of solitonic BSs~\cite{bezpalen,palenpani}. All these studies have in common that both stars are represented by the same complex scalar field, a feature typical for classical fluid stars. However, it might not be the most realistic model to describe unconnected Bose-Einstein condensates, where the complex scalar field could represent distinct quantized wave functions. Therefore, here we are interested in a different scenario where each star is described instead by an independent complex scalar field. This model generically represents Dark Stars (DS), which can be defined as any self-gravitating astrophysical compact objects which only interact through gravity with other stars. Notice that these regular objects behave as black holes, in the sense that they also interact only gravitationally, but present a wider range of possible compactness~\cite{Cardoso:2017njb}. Although DSs are modeled here with BSs, almost any kind of matter, i.e., either fermionic or bosonic, can be used to construct these objects, as far as the stars are represented with different matter species~\cite{Maselli2017}. It is important to stress that any binary formed by stars made with different non-interacting species would behave as Dark Stars --objects that only interact gravitationally--, like for instance the collisions of neutron and axion stars studied recently~\cite{2018arXiv180804668C}. Only the tidal properties of DSs might depend strongly on its composition.

In the present work, we aim to study the dynamics and the gravitational radiation produced during the coalescence of two DSs made by bosonic fields. These binary Dark Boson Stars (DBSs) consist on two BSs described by two different complex scalar fields, one for each star, satisfying the Einstein-Klein-Gordon (EKG) equations. Since the scalar fields of each star are different and there are no potential coupling them, there are only gravitational interactions between both DBSs.
Our simulations reveal not only the GW signature produced during the merger of DBS binaries as a function of the star's compactness, but also that the final remnant is always either a non-rotating superposition of independent BS (i.e., a multi-state BS~\cite{bernal}) or a spinning BH. Our results allow to extend our knowledge of BH waveforms to a wider range of compactnesses. They will also allow us to compare with standard binary BS~\cite{bezpalen,palenpani}, where there exists interaction between the stars both through gravity and matter. For our comparison purposes, we employ the potential and the same configurations
used in previous works~\cite{bezpalen,palenpani} for non-topological solitonic BS~\cite{frie,1992PhR...221..251L}, to construct our DBSs. 


This work is organized as follows. In Sec.~\ref{model} the model for DBSs binaries is introduced, describing equations of motion, numerical implementation, 
analysis quantities and the construction of initial data for both isolated and binary DBSs. In  Sec.~\ref{coales} we study numerically the coalescence of binary DBSs for different compactness, focusing on the dynamics and the gravitational radiation. Particular attention is paid to the comparison between DBSs and BSs binaries. In Sec.~\ref{conclu}, we summarize our results and present our conclusions. We have chosen geometric units such that $G=c=1$ and we adopt the convention where roman indices $a,b,c,\dots$ denote spacetime components (i.e., from $0$ to $3$), while $i,j,k,\ldots$ denote spatial ones (i.e., from $1$ to $3$).

\section{Model for Dark Boson Stars binaries }\label{model}

Our approach to model Dark Star binaries considers two independent complex scalar fields, interacting only through gravity, which are governed by Einstein-Klein-Gordon theory. Furthermore, a short description of the numerical implementation and analysis quantities is provided, together with a summary on the procedure to construct binary DBS initial data.

\subsection{Equations of Motion}

An arbitrary number of DBSs can be modeled by using a collection of $N$ complex scalar field (i.e., one for each star) interacting only through gravity. The dynamics of such a system can be described by the following action~\footnote{Notice that, although the context is different, this action is the same that describes multi-state BS~\cite{bernal}, where the super-index $(i)$ would denote each state of the scalar field.}
\begin{eqnarray}
S = \int d^{4}x\sqrt{-g}\left(\frac{R}{16\,\pi} - \sum_{i=1}^{N}\left[g^{ab}\nabla_{a}\bar{\Phi}^{(i)}\nabla_{b}\Phi^{(i)} \right. \right.\nonumber \\
\left. \left.  -\,\, V^{(i)}(|\Phi^{(i)}|^{2})\right] \frac{}{} \right)~~,
\label{actionMSBS}
\end{eqnarray}
where $R$ is the Ricci scalar associated to the metric $g_{ab}$ with determinant $g$. There are $N$ minimally coupled complex scalar fields $\Phi^{(i)}$, being $\bar{\Phi}^{(i)}$ their complex conjugate and $V^{(i)}(|\Phi^{(i)}|^2)$ the  scalar field potential.

We shall consider $N=2$ in order to study binary systems. The Einstein-Klein-Gordon evolution equations are obtained by taking the variation of the action~\eqref{actionMSBS} with respect to the metric $g_{ab}$ and each scalar field $\Phi^{(i)}$, namely 
\begin{eqnarray}
   R_{ab} - \frac{1}{2}g_{ab} R &=& 8\pi \, T_{ab} ~,\label{EKG1} \\
   g^{ab} \nabla_a \nabla_b \Phi^{(i)} &=& \frac{dV^{(i)}}{d \left|\Phi^{(i)}\right|^2} \Phi^{(i)}~,\label{EKG2}
\label{EKG3}
\end{eqnarray}
where $T_{ab}$ is the total scalar stress-energy tensor, given by the sum of the stress-energy tensors associated to each scalar field, 
\begin{eqnarray}
T_{ab} &=& T^{(1)}_{ab} + T^{(2)}_{ab}~~, \\
T^{(i)}_{ab} &=& \nabla_a \Phi^{(i)} \nabla_b \bar{\Phi}^{(i)} +    \nabla_a \bar{\Phi}^{(i)} \nabla_b \Phi^{(i)} \nonumber \\
&-&    g_{ab} \left[ \nabla^c \Phi^{(i)} \nabla_c \bar{\Phi}^{(i)}   + V^{(i)}\left(\left|\Phi^{(i)}\right|^2\right) \right].  
\end{eqnarray}  

Henceforward, we will consider a self-potential for each scalar field given by   
\begin{equation}
V^{(i)}(|\Phi^{(i)}|^2)=m_b^{2}\left|\Phi^{(i)}\right|^2\left(1 - \frac{2\left|\Phi^{(i)}\right|^{2}}{\sigma_{0}^{2}}\right)^2~,  
\end{equation}
where $m_{b}$ is related to the scalar field mass and $\sigma_{0}$ is a constant setting the compactness of the star. This kind of potential yields to non-topological solitonic boson stars~\cite{frie,1992PhR...221..251L}, which might have a compactness comparable or even higher than neutron stars~\cite{palenpani}. Notice that, with this choice for the potential, each scalar field is explicitly decoupled from the others. Therefore, the scalar field corresponding to each star interacts only with itself through its Klein-Gordon equations~\eqref{EKG2}, and with all the others through gravity by means of the spacetime metric described by Einstein equations~\eqref{EKG1}. This class of astrophysical compact objects behaves as stars made of dark matter (whence the name of Dark Boson Stars), in the sense that they  interact only gravitationally with other compact objects~\cite{Maselli2017}.

\subsection{Numerical implementation}

Einstein Equations can be written as a time evolution system by using the covariant conformal Z4 formulation
(CCZ4)~\cite{alic,bezpalen}, which is the one considered here.
The numerical discretization of these equations is performed by using the Method of Lines (MoL), which allows to separate time from spatial discretization. We use fourth-order accurate finite-difference operators for the spatial derivatives, together with a fourth order accurate Runge-Kutta time integrator. A sixth-order Kreiss-Oliguer dissipation is also included to eliminate unphysical high-frequency modes from our grid. The explicit form of the EKG evolution system and the discrete operators employed, together with numerical evolutions of binary BSs, can be found in Ref.~\cite{bezpalen,palenpani}.

These equations have been introduced in the platform {\it Simflowny}~\cite{simflowny_webpage,Arbona:2013,simf2,simf3} to automatically generate parallel code for the SAMRAI infrastructure~\cite{Hornung:2002,Gunney:2016,samrai_webpage}. SAMRAI provides parallelization and  Adaptive Mesh Refinement (AMR), which are crucial to obtain accurate solutions in an efficient manner by adding more resolution only where it is required (i.e., in the regions encompassing each BSs). We have also implemented, through {\it Simflowny}, an improved treatment of artificial AMR boundaries when there is sub-cycling in time~\cite{McCorquodale:2011,Mongwane:2015}.
Further details of the numerical implementation and convergence tests performed with this new platform can be found in Ref.~\cite{simf3}. 

Our simulations use Courant factors $\lambda_{c}\in \{0.357,0.15\},$  such that $\Delta t_{l} = \lambda_c \, \Delta x_{l}$ on each refinement level $l$ to guarantee that the Courant-Friedrichs-Levy condition is satisfied.
We use a domain $[-280,280]^3$ with 7 levels of refinement, each one with twice the resolution of the previous one, such that $\Delta x_{0}=4$ on the coarsest grid and $\Delta x_{6}=0.0625$ on the finest one.



\subsection{Analysis quantities}

Several quantities have been computed in order to analyze the dynamics of binary DBSs during their coalescence. One of them is the Noether charge, which is the analogous of the baryonic density in a fluid. For each scalar field, as a result of $U(1)$ invariance of the action~\eqref{actionMSBS}, there exists a Noether charge current defined by:
\begin{equation}
  J^{(i)\,a} = i g^{ab} (\bar{\Phi}^{(i)}\,\nabla_{b}\Phi^{(i)} - \Phi^{(i)}\,\nabla_{b} \bar{\Phi}^{(i)}),
\end{equation}
which implies a conserved quantity given by
\begin{equation}
N^{(i)} = \int d^{3}x\sqrt{-g}\,J^{(i)\,0}.
\label{noech}
\end{equation}
Notice that the conserved Noether charge $N^{(i)}$ can be interpreted as the number of bosonic particles in each star~\cite{liebpa}. Therefore, the total Noether charge of the system is given by the sum of the Noether charges of each star. 

The gravitational radiation can be described in terms of the Newman-Penrose scalar $\Psi_{4}$, which can be expanded in terms of spin-weighted $s=-2$ spherical harmonics~\cite{rezbish,brugman}, namely 
\begin{equation}
 r \Psi_4 (t,r,\theta,\phi) = \sum_{l,m} \Psi_4^{l,m} (t,r) \, {}^{-2}Y_{l,m} (\theta,\phi).
\label{eq:psi4}
\end{equation}
The instantaneous angular frequency of the gravitational wave can now be calculated easily from : 
 \begin{eqnarray}
 f_{GW} = \frac{\omega_{GW}}{2\pi}~,~~~
 \omega_{GW} = -\frac{1}{2}\Im\left(\frac{\dot{\Psi}_4^{l,m}}{\Psi_4^{l,m}}\right).
 \end{eqnarray}
 
A more direct quantity, related directly to the response of the detector, is the strain defined as $h(t)=h_{+}(t) - i\,h_{\times}(t),$ where $(h_{+},h_{\times})$ are the plus and cross modes of gravitational waves. The Newman-Penrose scalar $\Psi_{4}$ is related to the strain via:
\begin{equation}
\Psi_{4} = \ddot{h}_{+} - i\,\ddot{h}_{\times},
\end{equation} 
The components of the strain in the time domain can be calculated by performing the inverse Fourier transform of the strain  in the frequency domain, $h^{l,m}(t) \equiv {\cal F}^{-1} [{\tilde h}^{l,m}(f)]$, which can be calculated as~\cite{2011CQGra..28s5015R} 
\begin{equation}
{\tilde h^{l,m}(f)} = 
\begin{cases} 
        \displaystyle - \frac{{\cal F} [{\Psi_4}^{l,m}(t)]}{f_{0}^2} ~, &  f < f_{0}\\\\
       \displaystyle - \frac{{\cal F} [{\Psi_4}^{l,m}(t)]}{f^2} ~, &  f\geq f_{0}    
\end{cases}\, ,
\end{equation}
where $f_{0}$ is the initial orbital frequency. 

During the evolution we compute also other global quantities integrated over a spherical surface, like the Arnowitt-Deser-Misner (ADM) mass and the angular momentum. These global quantities, together with $\Psi_4$, are calculated in spherical surfaces at different extraction radii, although we only show the results obtained at $R_{ext}=50$.

\subsection{Initial data}


{\bf Single Dark BS.} Initial data for isolated DBSs is exactly the same as for BSs. The method to construct these solutions can be summarized as follows:
\begin{itemize}
	\item consider a spherically symmetric static spacetime in Schwarzschild-like coordinates $\tilde{r}$, together with a complex scalar field with an harmonic ansatz, namely:
\begin{eqnarray}\label{idbs_metric}
ds^{2} &=& -\alpha^{2}(\tilde{r})dt^{2}+a^{2}(\tilde{r})d\tilde{r}^{2}+\tilde{r}^{2}d\Omega^{2}, \\
\label{idbs_sf}
\Phi(t,\tilde{r}) &=& \phi(\tilde{r})\,e^{-i\omega\,t},
\end{eqnarray}

\item the EKG system~\eqref{EKG1}-\eqref{EKG3}, with the previous assumptions ~\eqref{idbs_metric}-\eqref{idbs_sf}, reduces to a set of ordinary differential equations (ODE). These ODEs  can be integrated by imposing suitable boundary conditions (i.e., regularity at the origin and an asymptotic flatness at infinity).

\item transform the solution from Schwarzschild-like coordinates into isotropic coordinates $r$,
\begin{eqnarray}
ds^{2} = -\alpha^{2}(r) + \psi^{4}(r)(dr^{2} + r^{2}d\Omega^{2}),
\end{eqnarray}
 by performing a numerical integration. The transformation from these coordinates to Cartesian ones is trivial. Further details of this procedure can be found in Ref.~\cite{2004PhDT.......230L}.
\end{itemize}

The solution is a Boson Star, a regular self-gravitating compact object made of a complex scalar field. In contrast to NSs, BSs does not present a hard surface that woud allow to define unambiguously their size. Nevertheless, one can define the radius containing $99\%$ of the ADM mass (or the Noether charge), denoted as $R_{M}$ (or $R_{N}$). Notice that, in these spherically symmetric solutions, the ADM mass can be easily computed as:
\begin{equation}
M_{ADM} = \lim\limits_{r\to\infty}\frac{r}{2}\left(1 - \frac{1}{\alpha(r)^{2}}\right).
\end{equation} 
This way, one can define the compactness of a BS as $C=M_{ADM}/R_{M}$. Notice that this quantity ranges between $C\approx 0.1-0.25$ for NSs and between $C=0.5-1$ for BHs (non-rotating and with extremal spin, respectively).

\begin{figure}
\centering
\includegraphics[width=0.485\textwidth]{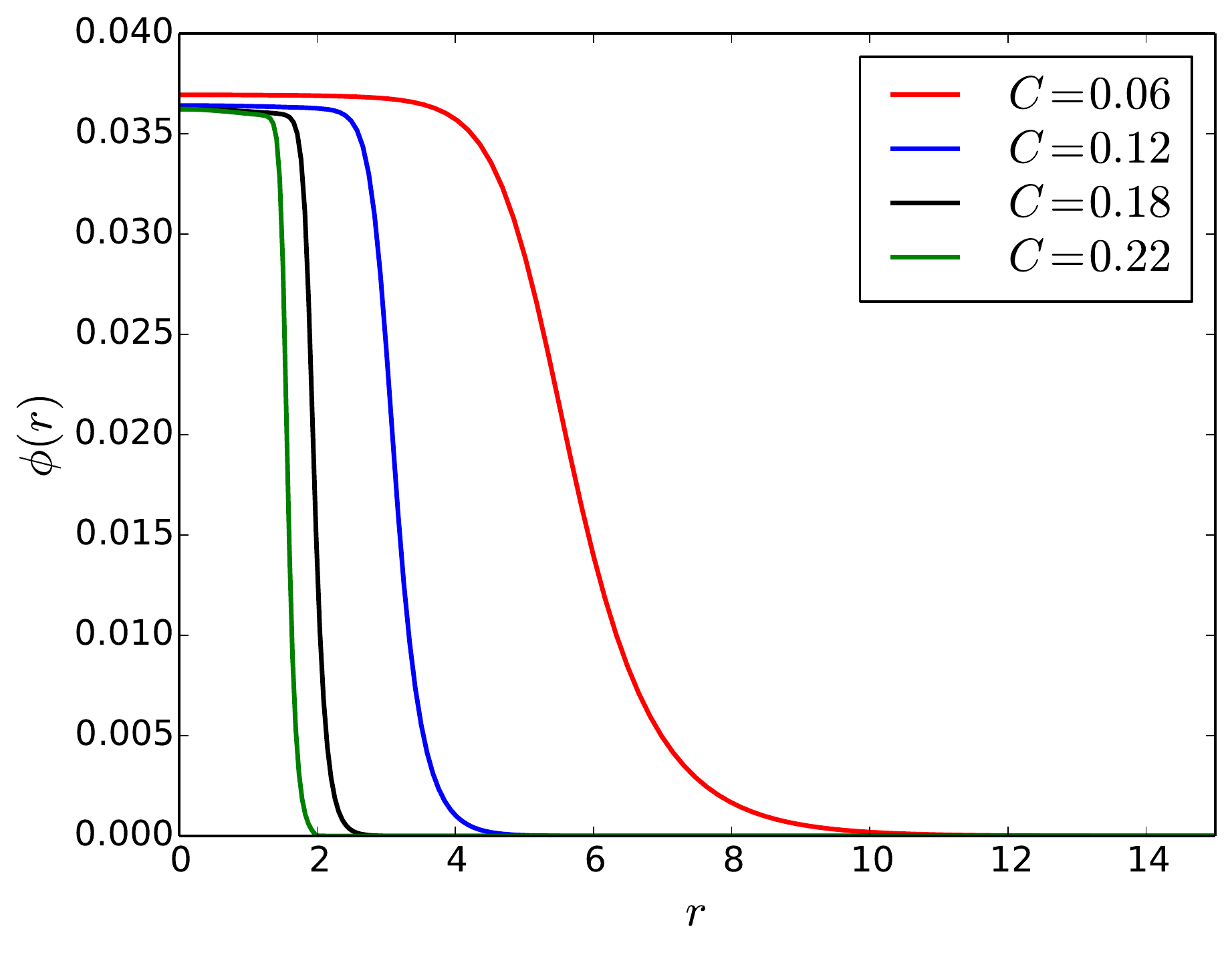}
\caption{{\em Initial data of DBS}. Radial profile of the scalar field $\phi(r)$ for each compactness. Notice that it is nearly constant in the interior and then falls off exponentially at the surface of the star. This fall off is steeper as the compactness increases.}   
\label{profiles}
\end{figure}

\begin{figure}
\centering
\includegraphics[width=0.48\textwidth]{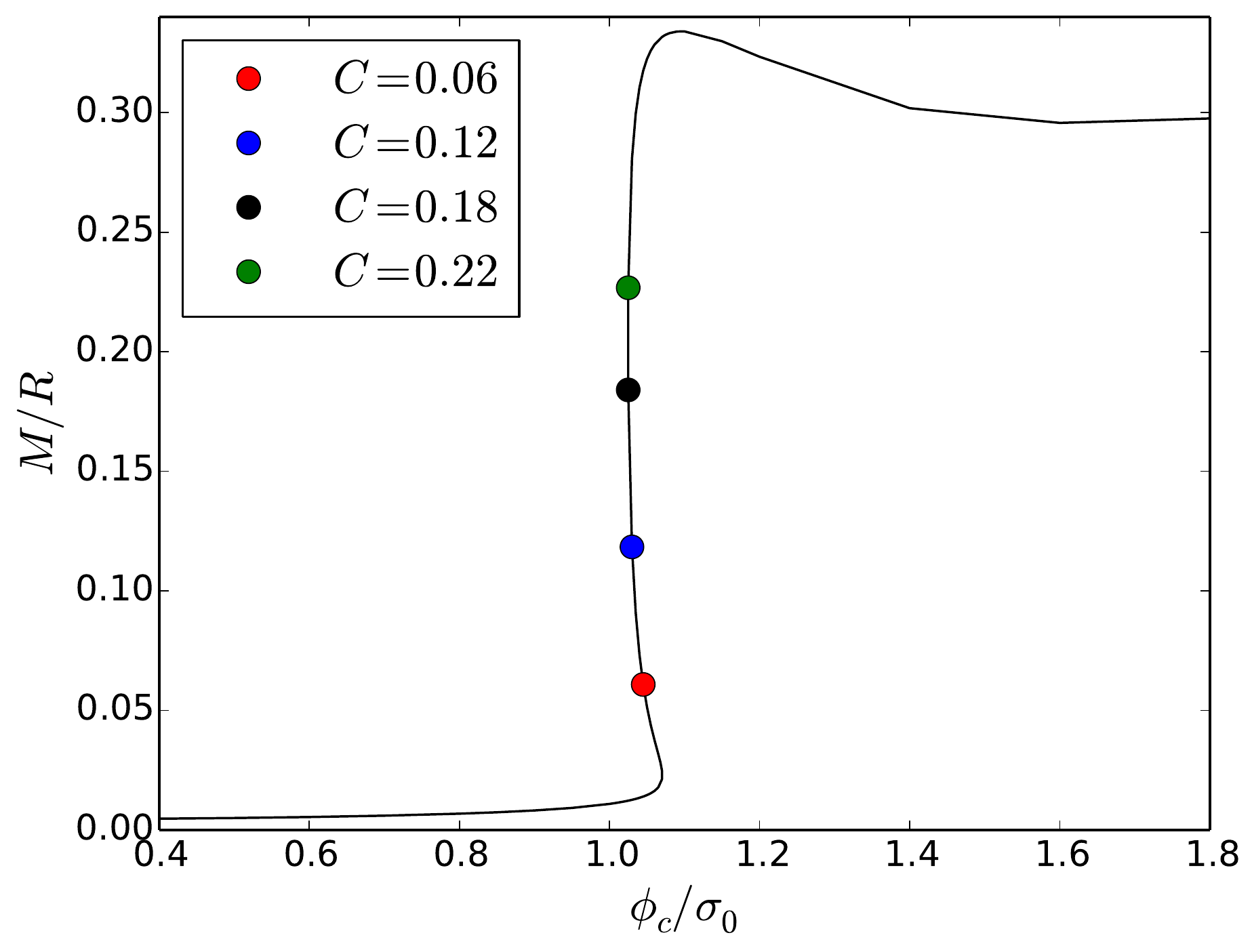}
\caption{ {\em Initial data of DBS}. Compactness as a function of the central value of the scalar field $\phi_{c}$ with $\sigma_0=0.05$. Circular markers refer to the \emph{initial} equilibrium configurations considered both here and in previous work~\cite{palenpani} to construct initial data for binaries.}   
\label{compvsphi}
\end{figure}

In the present work, we consider the same configurations investigated in~\cite{palenpani}. Therefore, we restrict ourselves to the choice $\sigma_{0}=0.05$, which lead to highly compact BS, and construct four stars with compactness $C=\{0.06,0.12,0.18,0.22\}$ belonging to the stable branch (i.e., equilibrium configurations which are stable under small perturbations). The radial profile of the scalar field for each compactness is displayed in Fig.~\ref{profiles}. As it is shown in Fig.~\ref{compvsphi}, all these initial configurations are well inside the stable branch, which is the curve on the left of the maximum compactness $C_{max}=0.33$. 

Each star has been rescaled, by a suitable choice of $m_b$, such that $M_{ADM}=0.5$. In Table~\ref{isolateddbs} can be found all the characteristics of our isolated BSs models: central value of the scalar field $\phi_c$, bare mass of the boson $m_b$, Noether charge $N$, radius of the star $R_M$ and $R_N$, angular frequency of the phase of $\phi$ in the complex plane $\omega$ and dimensionless tidal Love number $k_{tidal}$~\cite{2017arXiv170101116C,Sennett:2017etc}. As a reference, notice that $k_{\rm tidal}\approx 200$ for a neutron star with a typical equation of state, while that $k_{\rm tidal}=0$ for BHs.

\begin{table}
\begin{ruledtabular}
\begin{tabular}{c||cc|ccc|c}
 $C$ & $\phi_c/\sigma_0$ &  $m_b$  & ${N}$ & $(R_M,R_N)$ & ${\omega}$ & $k_{\rm tidal}$ 
\\ \hline\hline 
 0.06 & 1.045 &  0.9880   &  0.9867  & (8.21,\,7.38)  & 0.3828   & 8420\\ 
 0.12 & 1.030 &  2.9124   &  0.4785  & (4.22,\,3.88)  & 0.7787   & 332  \\ 
 0.18 & 1.025 &  6.2514   &  0.2929  & (2.71,\,2.54)  & 1.2386   & 41 \\ 
 0.22 & 1.025 &  8.5663   &  0.2417  & (2.31,\,2.19)  & 1.4725   & 20
\end{tabular}
\caption{{\em Characteristics of isolated BSs models with $\sigma_0=0.05$.}  The table shows, in units such that $M_{ADM}=0.5$, the following quantities: compactness, central value of the scalar field, bare mass, Noether charge, radius of the star (i.e, containing $99\%$ of either the mass, $R_M$, or the Noether charge, $R_N$), angular frequency of the phase of $\phi$ in the complex plane and dimensionless tidal Love number ($k_{\rm tidal}$).
}\label{isolateddbs}
\end{ruledtabular}
\end{table}

{\bf Binary Dark BS.} Initial data for  binary BS can be constructed by using a superposition of two boosted isolated BS solutions. Since we are interested on modeling DBS binary systems, which only interact through gravity, we proceed as follows:
\begin{itemize}
	\item the solution of each BS, calculated as described in the previous subsection, is extended to Cartesian coordinates $\{g^{(i)}_{ab}(x,y,z),\Phi^{(i)}(t,x,y,z)\}$.
	
	\item the spacetime of binary DBS is obtained by a superposition of the isolated spacetimes of two BSs, centered at positions $(0,\pm y_{c},0)$ and with a boost $\pm v_x$ along the $x$-direction. The scalar field of each boosted star is not modified by the other star. The full solution can be expressed then as:
	\begin{eqnarray}
	g_{ab} &=& g^{(1)}_{ab}(x,y-y_{c},z;+v_{x})\nonumber \\
	& & +\,\, g^{(2)}_{ab}(x,y+y_{c},z;-v_{x}) - \eta_{ab},\\
	\Phi^{(1)} &=& \Phi^{(1)}(x,y-y_{c},z;+v_{x}),\\
	\Phi^{(2)} &=& \Phi^{(2)}(x,y+y_{c},z;-v_{x}),
	\end{eqnarray}
	where $\eta_{ab}$ is the Minkowski metric in Cartesian coordinates.
\end{itemize}

Notice that a fine-tuning of the initial orbital velocity is required to set the binary in a quasi-circular orbit. It is also worthwhile to emphasize that this superposition does not satisfy the energy and momentum constraints due to the non-linear character of Einstein's equations. However, our evolution formalism enforces dynamically an exponential decay of these constraint violations (for instance, see Fig.\,10 in Ref.~\cite{bezpalen}). Nonetheless, convergence tests performed on the most stringent case indicates that our initial data is accurate enough to allow us investigate the problem at hand.

\section{Coalescence of Dark Boson Stars }\label{coales}

The coalescence of binary identical DBSs obtained from numerical simulations is analyzed in detailed, focusing on the dynamics and the gravitational radiation produced with different star's compactness. Furthermore, in order to infer the effect of  matter interactions, these results for dark BS binaries (i.e., only gravity interactions) are contrasted with those for standard BS binaries (i.e., both gravity and matter interactions). Since the individual mass of each star in isolation is $M=0.5$,  the binary has approximately a total initial mass $M_0\approx 1$. 

\subsection{Dynamics}

First of all, let us mention that the initial boost velocities, required to set the binary system roughly in a quasi-circular orbit, are different for each case due to several reasons. The largest difference appears for the case with $C=0.06$ because the radius of the stars (i.e., and, consequently, their initial separation) is significantly larger than in the other cases, implying a lower boost velocity. Since the separation in all the other binary configurations is the same, discrepancies on the boost velocities arise from slight differences on the star's  compactness and total mass of the system. More detailed information regarding the initial parameters is available in Table~\ref{table2}, together with the main properties of each binary DBS and its final remnant.
   
\begin{figure*}
\centering
\includegraphics[width=17cm, height=15cm]{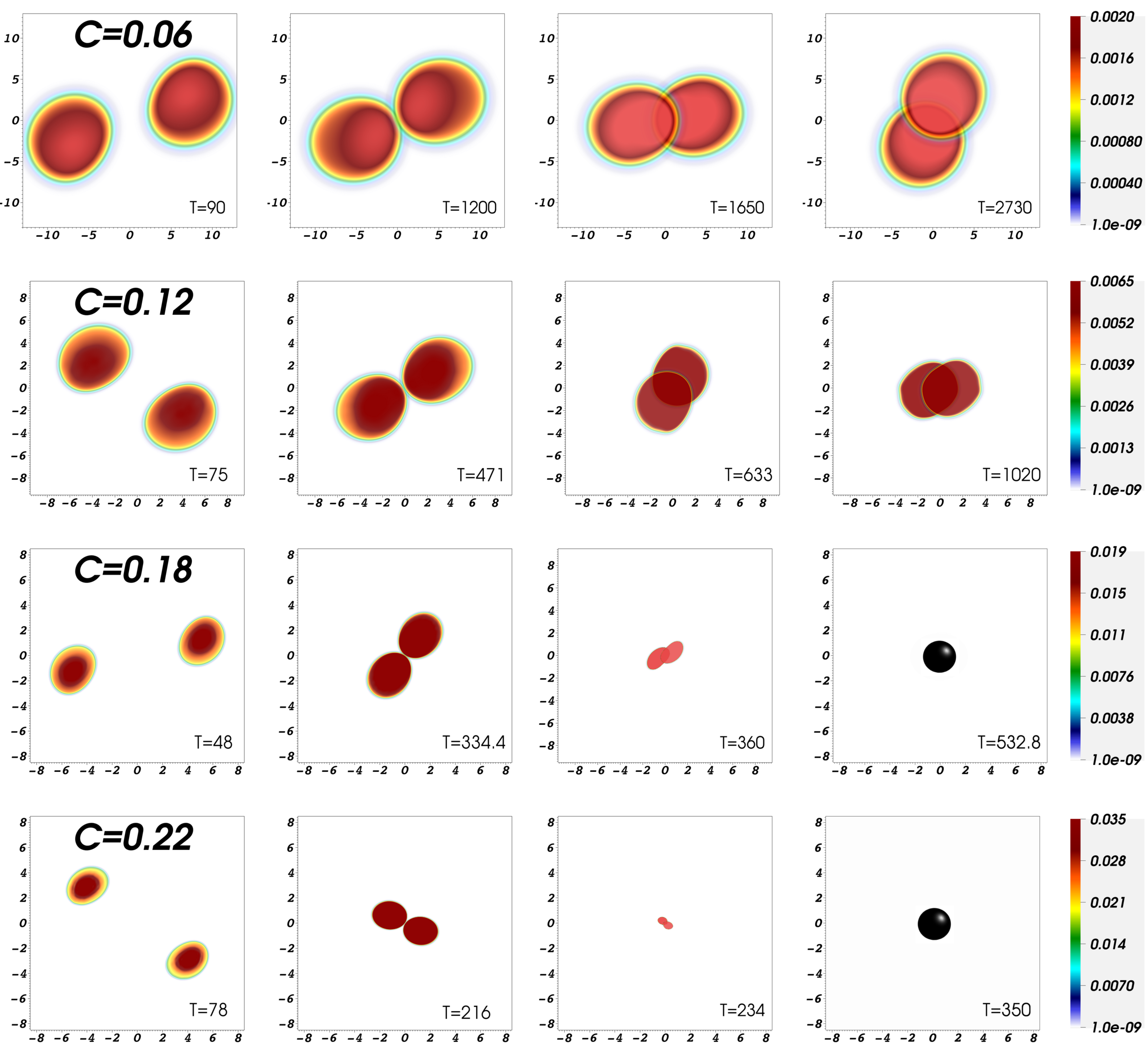}
\caption{{\em Dynamics of DBS coalescence}. Noether charge densities, corresponding to the individual stars, in the equatorial plane at several illustrative times. Each row corresponds to a different star's compactness (from top to bottom, $0.06$, $0.12$, $0.18$, and $0.22$). First column illustrates a time in the early inspiral, the second one is roughly at contact time, the third one is during the merger stage and the fourth one at the end of our simulation. Notice that the final remnant for $C \lesssim 0.12$ is composed by two rotating co-existing DBS, while that for $C \gtrsim 0.18$ is
a rotating BH (i.e., the black sphere at late times represents
the apparent horizon).}
\label{noethers}
\end{figure*}

\begin{figure*}
\centering
\includegraphics[width=17cm, height=15cm]{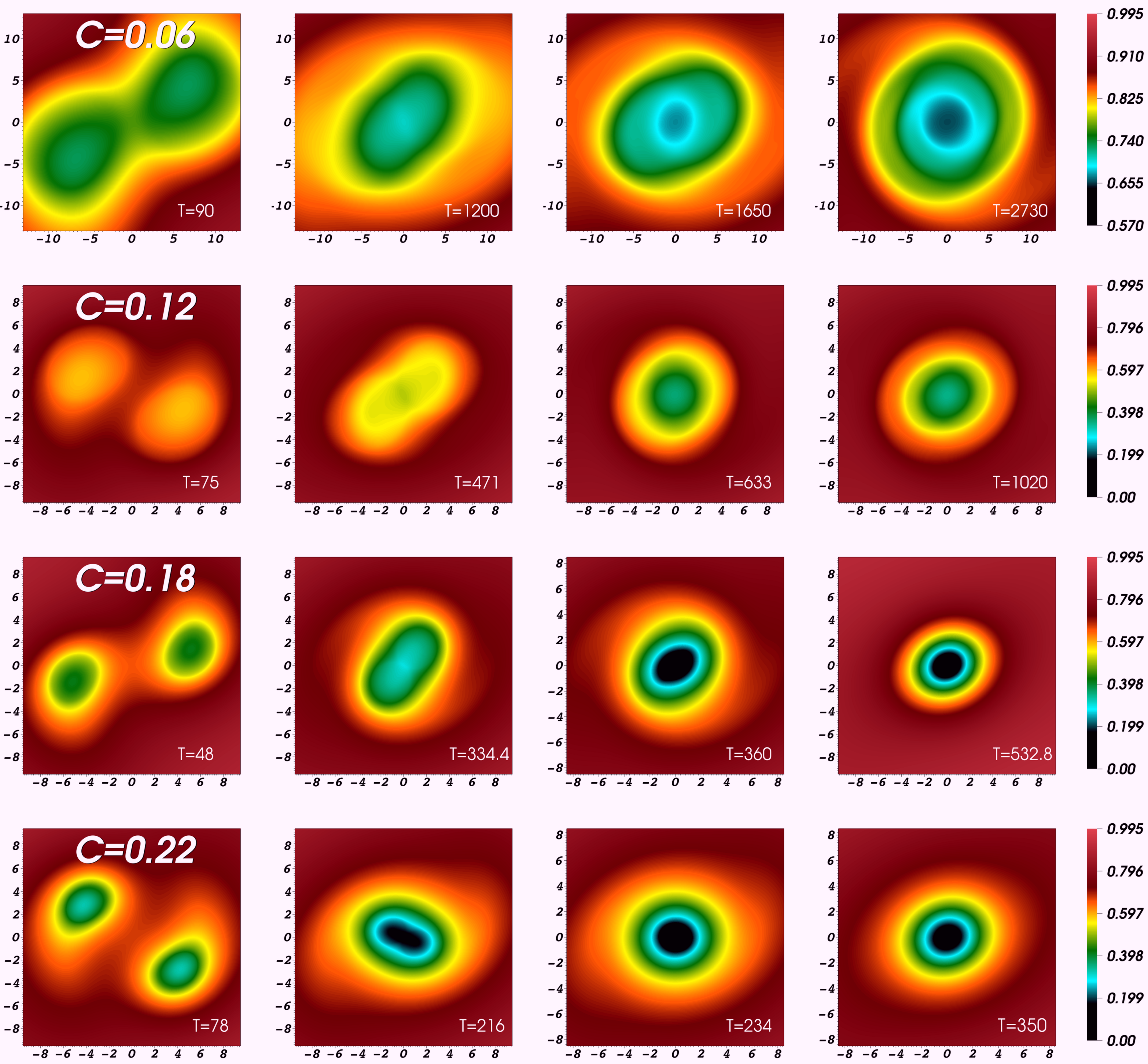}
\caption{{\em Dynamics of DBS coalescence}. Conformal factor, which gives a rough description of the gravitational potential, in the equatorial plane at the same time snapshots as in Fig.~\ref{noethers}.  Each row corresponds to a different compactness (from top to bottom, $0.06$, $0.12$, $0.18$, and $0.22$). }
\label{chi}
\end{figure*}

Some snapshots of the Noether charge density for each case, at different representative times of the coalescence, are displayed in Fig.~\ref{noethers}. The conformal factor of the metric used in the CCZ4 formalism, which represents roughly the gravitational potential, is displayed in Fig~\ref{chi} at the same times. As stated above, the interaction between DBSs takes place only through gravity, even when there exists an overlap between the stars, as shown in Fig.~\ref{noethers}.
Therefore, the first feature that one can observe is that the transition between inspiral and merger stages depends on the star's compactness. For low compact stars $C \lesssim 0.12$, the inspiral phase does not finish suddenly at the contact time $t_{c}$ (i.e., defined as the time at which the individual Noether charge densities make contact for the first time) but smoothly continues to the merger phase. The final remnant is a superposition of two coexisting orbiting DBSs that, at late times, are expected to settle down into a stationary spherically-symmetric configuration equivalent to a multi-state BSs~\cite{bernal}.
For high compact stars $C \gtrsim 0.18$, the transition is quite abrupt and clearly distinguishable. The final remnant is too compact and inevitably collapses to a rotating BH.
Therefore, the final object also depends on the initial compactness of the identical stars. 
We can identify a critical transition compactness $C_{T}$ such that below
that value the remnant relaxes to a multi-state BS and above it collapses to a BH. From our simulations we can infer that $0.12 < C_{T} < 0.18.$

\begin{figure}
\centering
\includegraphics[width=1.0\linewidth]{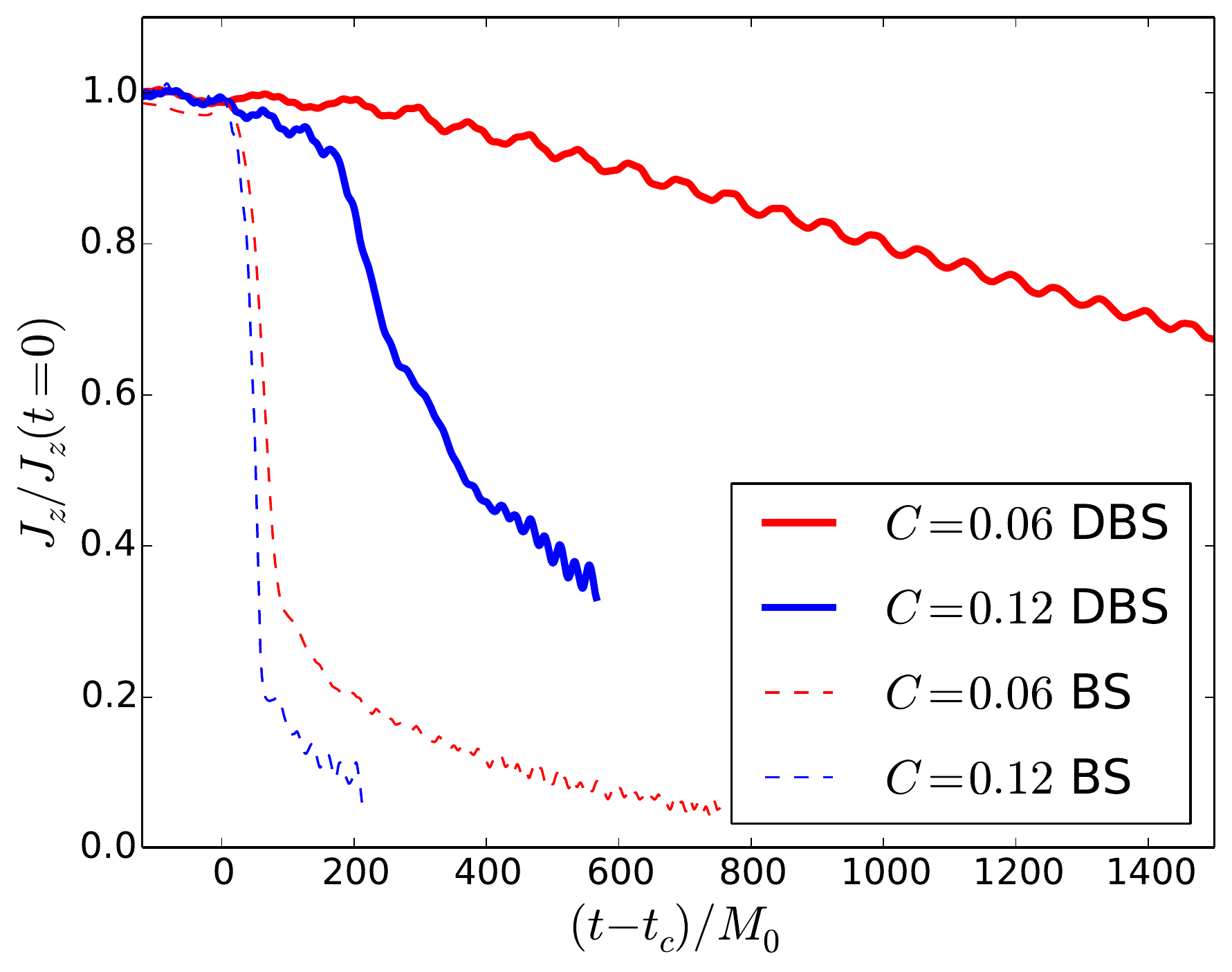}
\caption{ {\em Dynamics of DBS coalescence}. Angular momentum $J_{z}$ as a function of time for DBS with initial compactness $C=0.06$ and $C=0.12$. This quantity, for the remnants of DBS mergers, decays to zero in a much longer timescale than those of BS mergers, especially for the lowest compactness $C=0.06$. Notice that the sudden decay of the binary BS case with $C=012$ was enhanced by the ejection of two blobs of scalar field during the merger~\cite{palenpani}.}
\label{analysis}
\end{figure}

As discussed before, the most compact cases lead to the formation of a rotating BH, a process that is common to other binary mergers. Then, we shall focus our analysis to the more peculiar and distinctive scenario where the merger does not produce a BH. Instead, the two stars keep rotating around each other while emitting GWs, despite a significant overlap between the individual Noether charge densities. In contrast to the inspiral phase, after the contact time the two stars can not be modeled as point sources, since the distance between their centers of mass is comparable or smaller than their radius. 
The evolution of angular momentum is displayed in Fig.~\ref{analysis} for these low-compact cases.  Angular momentum is radiated slowly during the coalescence through gravitational waves, increasing rapidly its emission rate after the contact time. Meanwhile, the mass only decreases by roughly $10\%$ at most, which means that the final multi-state BS has approximately the total initial mass of the binary system. For comparison purposes, the angular momentum evolution of the corresponding binary BSs have been added to the same plot. The  behavior of BS after the contact time is clearly different, showing a sharp decay due to the interaction between the scalar fields modeling each star. A deeper discussion of the differences between DBS and BS will be given in Section~\ref{comp}.    
 
\begin{table*}
\begin{ruledtabular}
\begin{tabular}{c||cccc|c|ccc|ccc}
 $C$ & $y_c^{(i)}$ & $v_x^{(i)}$ & $M_0$ & $J_0$  & 
   $t_\mathrm{c}$ &
   remnant & $E^{DBS}_\mathrm{rad}/M_0$  &  ${\cal E}^{DBS}_\mathrm{rad}/M_0$ &
   remnant & $E^{BS}_\mathrm{rad}/M_0$ &  ${\cal E}^{BS}_\mathrm{rad}/M_0$\\
 \hline\hline 
 0.06 & $\pm 8$ & $\pm 0.142$ & 1.07  & 1.16   &
  1200  & BS+BS & 0.068 & 0.06 & BS & 0.075 & 0.029\\
 0.12 & $\pm 5$ & $\pm 0.210$ & 1.18  & 1.24  &
 471  &  BS+BS & 0.127 & 0.12 & BS & 0.085 & 0.057\\ 
 0.18 & $\pm 5$ & $\pm 0.214$ & 1.29  & 1.40    & 
 335  &  BH & 0.014 & 0.18 & BS & 0.120 & 0.086 \\ 
 0.22 &  $\pm 5$ & $\pm 0.220$ & 1.46  & 1.65 & 
 218  & BH & 0.030 & 0.22 & BH &  0.030 & 0.1
\end{tabular}
\caption{ {\em Characteristics of binary of DBS models.}
The entries of the table are, respectively: the compactness $C$ of the individual DBSs in the binary, the initial positions $y_c^{(i)}$, the initial velocities of the boost $v_x^{(i)}$, the initial total ADM mass $M_0$, the initial total orbital angular momentum $J_0$ of the system, the time of contact of the two stars $t_c$,  the final remnant, the total radiated energy in gravitational waves for each simulation $E_\mathrm{rad}$ (i.e., integrated from the beginning and extrapolated to large times after the contact time) and the one estimated analytically ${\cal E}_\mathrm{rad}$ as described in Appendix~\ref{appA}. We also included previous results corresponding to binary BSs for comparison purposes.
}
\label{table2}
\end{ruledtabular}
\end{table*}

\subsection{Gravitational Radiation}

\begin{figure}
\centering
\includegraphics[width=1.0\linewidth]{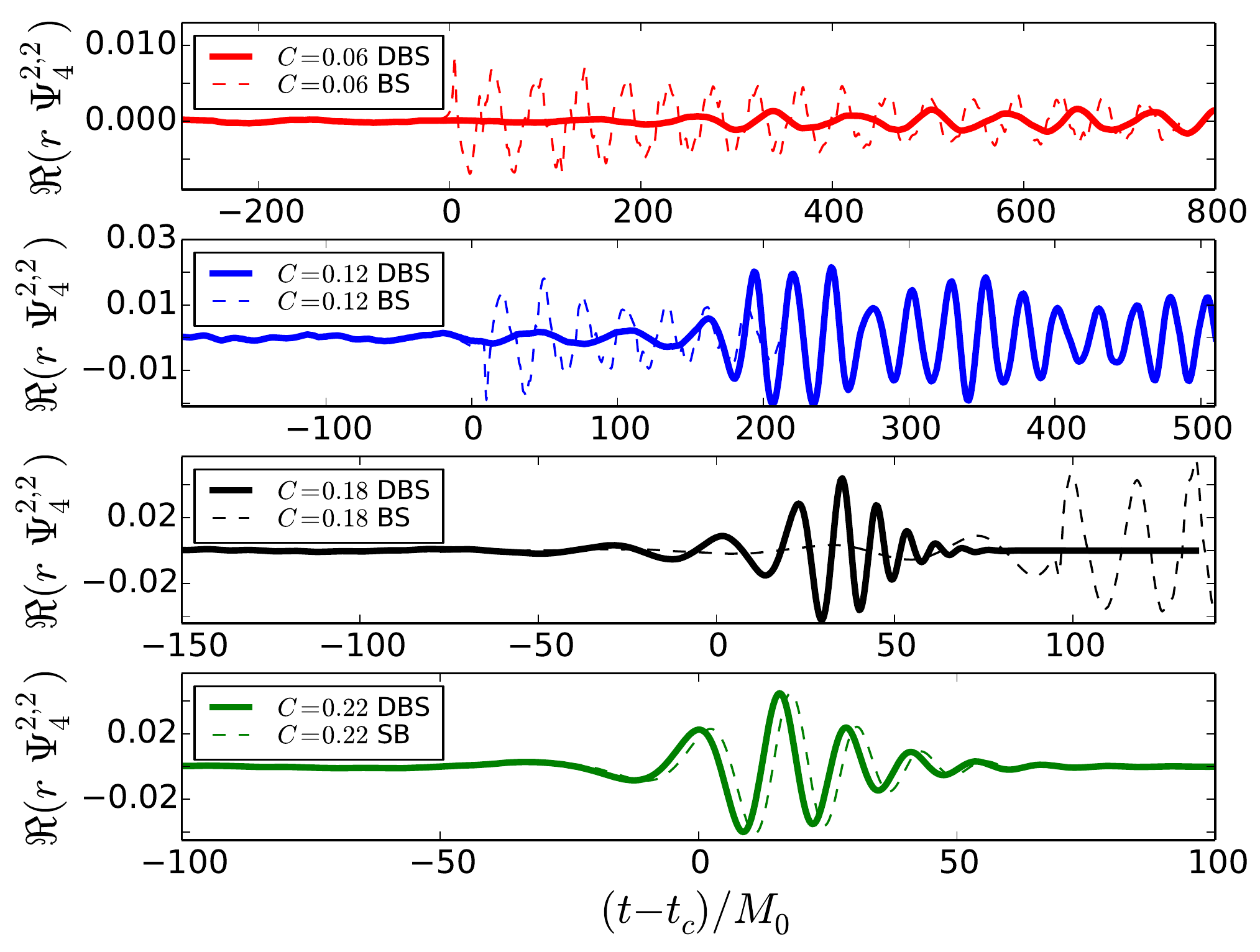}
\caption{ {\em Gravitational waves}. The real part of the main $l=m=2$ mode of $\Psi_4$ describing the gravitational emission produced by DBS and BSs binaries as a function of time.}
\label{psi4}
\end{figure}

\begin{figure*}
\centering
\includegraphics[width=1.0\linewidth]{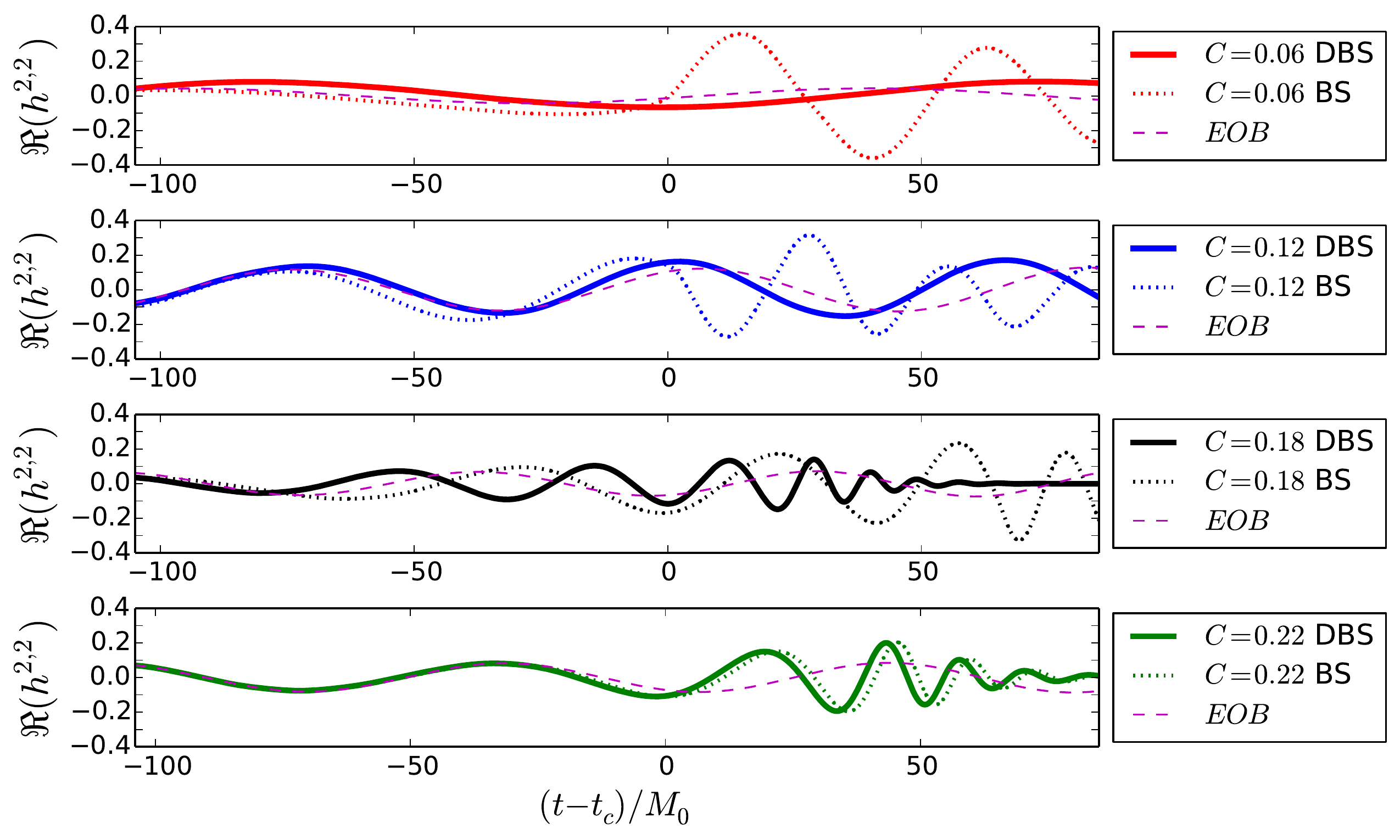}
\caption{ {\em Gravitational waves}. Main mode of the strain for DBS binaries with different compactness near the contact time. All cases are compared to the EOB approximation of a quasi-circular binary BH coalescence~\cite{2017PhRvD..95d4028B} by matching the waveforms at early inspiral.} 
\label{gwstrain}
\end{figure*}

The main mode $l=|m|=2$ of the Newman-Penrose scalar $\Psi_{4}$, encoding the gravitational radiation produced during the coalescence, is displayed in Fig.~\ref{psi4}. Furthermore, the same mode of the strain, near the contact time, is shown in Fig.~\ref{gwstrain}. Notice that time has been rescaled with the initial total mass $M_0$ of each binary, and shifted such that contact time occurs at $t=0$. 

Let us start with the less compact cases, $C \lesssim 0.12$, whose merger leads to a superposition of two co-existing BSs as a final state. As it might be expected, gravitational radiation produced during the early inspiral is weak, since the stars have large radii and therefore collide at a low frequency. After making contact, the binary enters smoothly to the merger stage, with both stars orbiting around each other for a long time and yielding to stronger gravitational waves than during the inspiral. Losses of energy and angular momentum occur mainly during this stage, while the remnant formed by the two coexisting stars rotates at a faster frequency, radiating more intense GWs. Although the cases with $C \gtrsim 0.18$ share the same behavior in the early inspiral, soon after the contact time the remnant becomes unstable and collapses to a rotating BH. The exponential decay of the gravitational wave signal observed for these cases in Fig.~\ref{psi4} and Fig.~\ref{gwstrain} is a clear evidence on this final BH state.

Furthermore, we have included the comparison with an effective-one-body (EOB) approximation that describes the adiabatic coalescence of quasi-circular binary BHs ~\cite{2017PhRvD..95d4028B}. Interestingly, although the signals show a good agreement during the inspiral phase, they are quite different as they get closer to the contact time: analogous to NSs, binaries composed by DBSs are strongly affected by tidal interacting forces only when they are in a close orbit. This means that, even though DBSs 
only suffer gravitational interactions, they behave yet rather different than BHs.
The most compact case, $C=0.22$ shows that even when the final fate of the remnant is a BH, DBSs and BHs are still different in the late inspiral and merger. An accurate quantitative comparison of our simulations with the EOB approximation would require initial data for DBS binaries in quasi-circular orbits with much smaller constraint violation, which is not yet available.

\begin{figure*}
\centering
\includegraphics[width=1.0\linewidth]{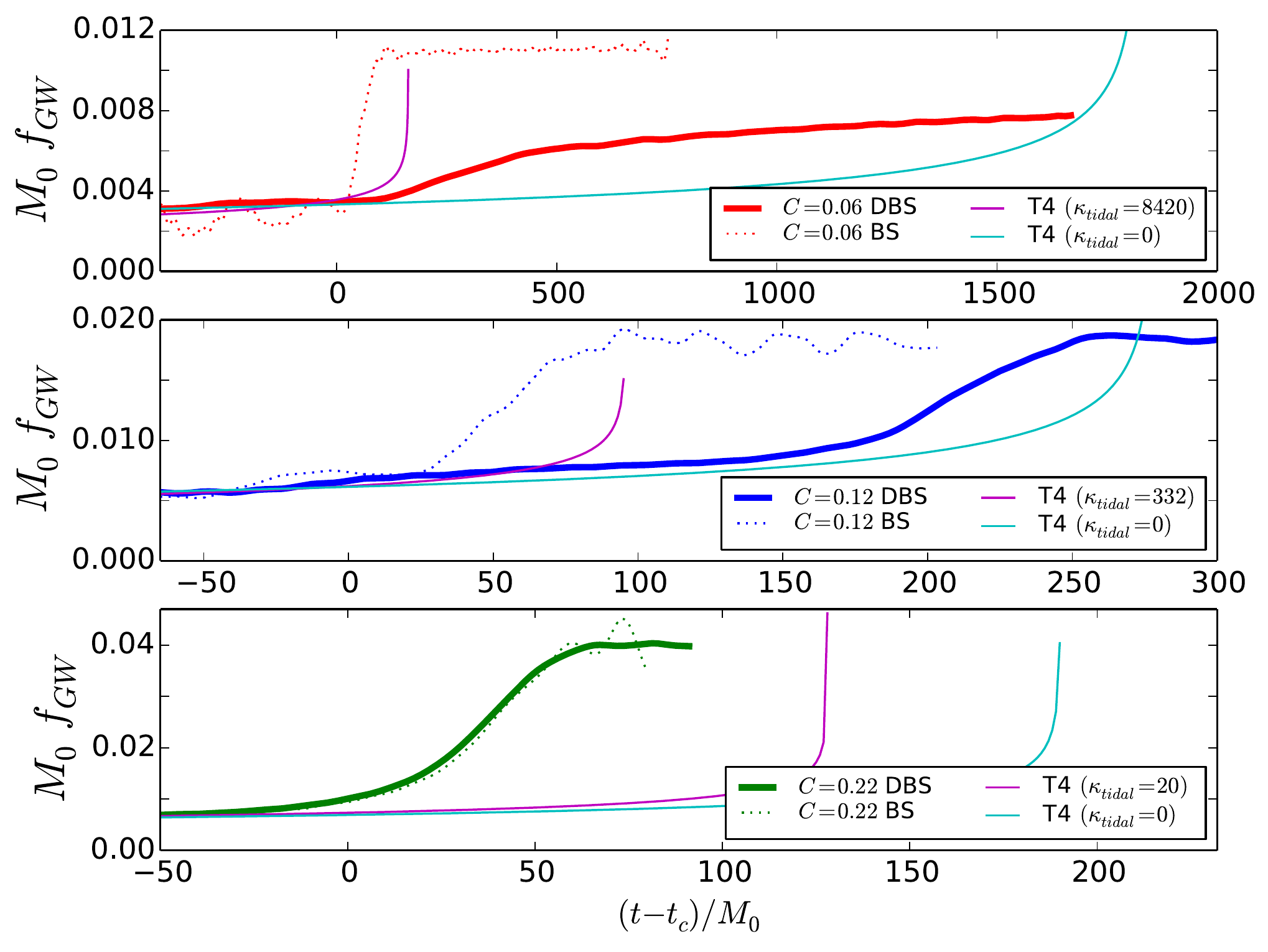}
\caption{ {\em Gravitational waves}. GW wave frequency $f_{GW}=\omega_{GW}/2\pi$ as a function of time, where $\omega_{GW}$ is the instantaneous GW angular frequency from the main $l=m=2$ mode. The frequencies calculated numerically for DBS and BS are displayed in thick solid and dashed lines, respectively, while that frequencies calculated by Taylor T4 approximation with and without tidal effects are plotted in thin solid lines. Notice that significant differences arise just after contact time.}
\label{PPcomp}
\end{figure*}

It is also illustrative to analyze the instantaneous GW frequencies $f_{GW}$ of DBS binaries and compared them to:  (i) BS binaries, (ii) a Post-Newtonian (PN) T4 approximation for point particles(i.e., BHs)~\cite{T4}, and (iii) a T4 approximation including also the lowest order tidal effects~\cite{T4tidal1,T4tidal2}, whose  strength can be measured by the tidal Love number~\cite{2017arXiv170101116C,Sennett:2017etc}.  Fig.~\ref{PPcomp} displays these four models (i.e., DBS, BS and T4 with and without tidal effects) for each compactness. Again, all models behave similarly during the inspiral, with differences arising near the contact time. While the DBS binary with $C \lesssim 0.12$  exhibits a smooth and soft increase on the frequency, the corresponding BS binary shows an abrupt rise as a consequence of the stronger dynamics of the remnant, induced by scalar field interactions. 
The PN-T4 approximation, either with or without tidal effects, leads to different frequencies than the DBS binary after the contact time, such that $f_{GW}$ of DBS is somewhere in between these two approximations. This means that, although including tidal effects at the lowest order might be acceptable during the inspiral phase, it is not accurate enough after the contact time when the effects of extended bodies become important.
Notice also that the frequencies calculated from T4-PN are closer to those of the BS binary when tidal effects are activated, probably because both matter interactions and tidal effects accelerate the dynamics of the system. 

The post-merger frequency of DBS and BS reach roughly the same value in the case $C=0.12$. We presume that in the case $C=0.06$ both models will also reach the same frequency at the end state, although the time scales for the remnant to settle down are much longer than considered on this work. In the most compact cases, $C= 0.22$, there are no significant differences between BS and DBS frequencies, as we have seen in the waveform as well. These high-compact cases are noticeably different than the T4 approximations (i.e., with and without tidal effects), showing again that such merger is still different from the one of a binary BH.  
Notice that the case $C=0.18$ is not directly comparable since the remnants of binary DBS and BS are different. 


\begin{figure}
\centering
\includegraphics[width=1.0\linewidth]{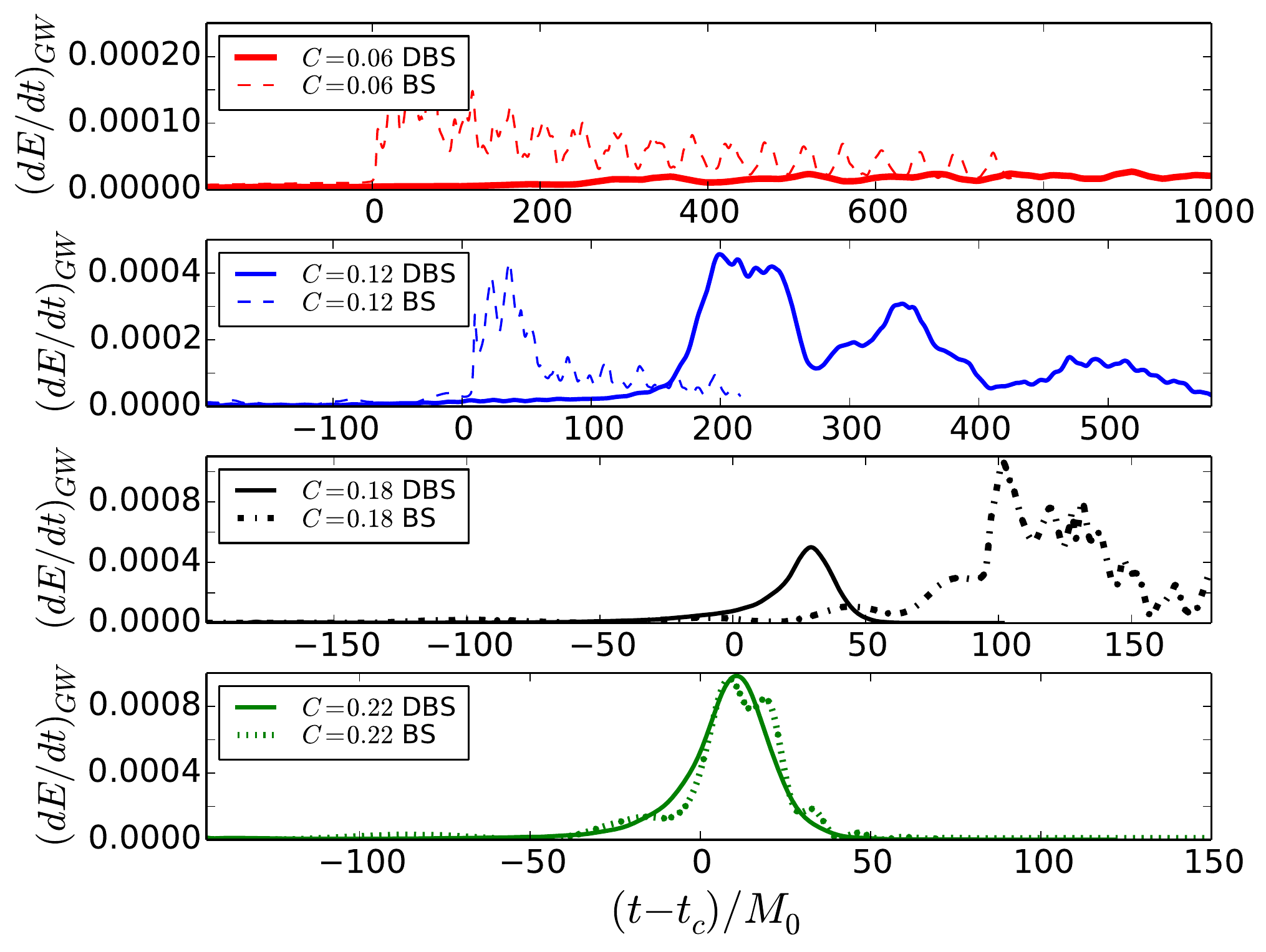}
\caption{ {\em Gravitational waves}. Luminosity of gravitational waves radiated during the coalescence of DBS and BS binaries.}
\label{lum2}
\end{figure}

\begin{figure}
\centering
\includegraphics[width=1.0\linewidth]{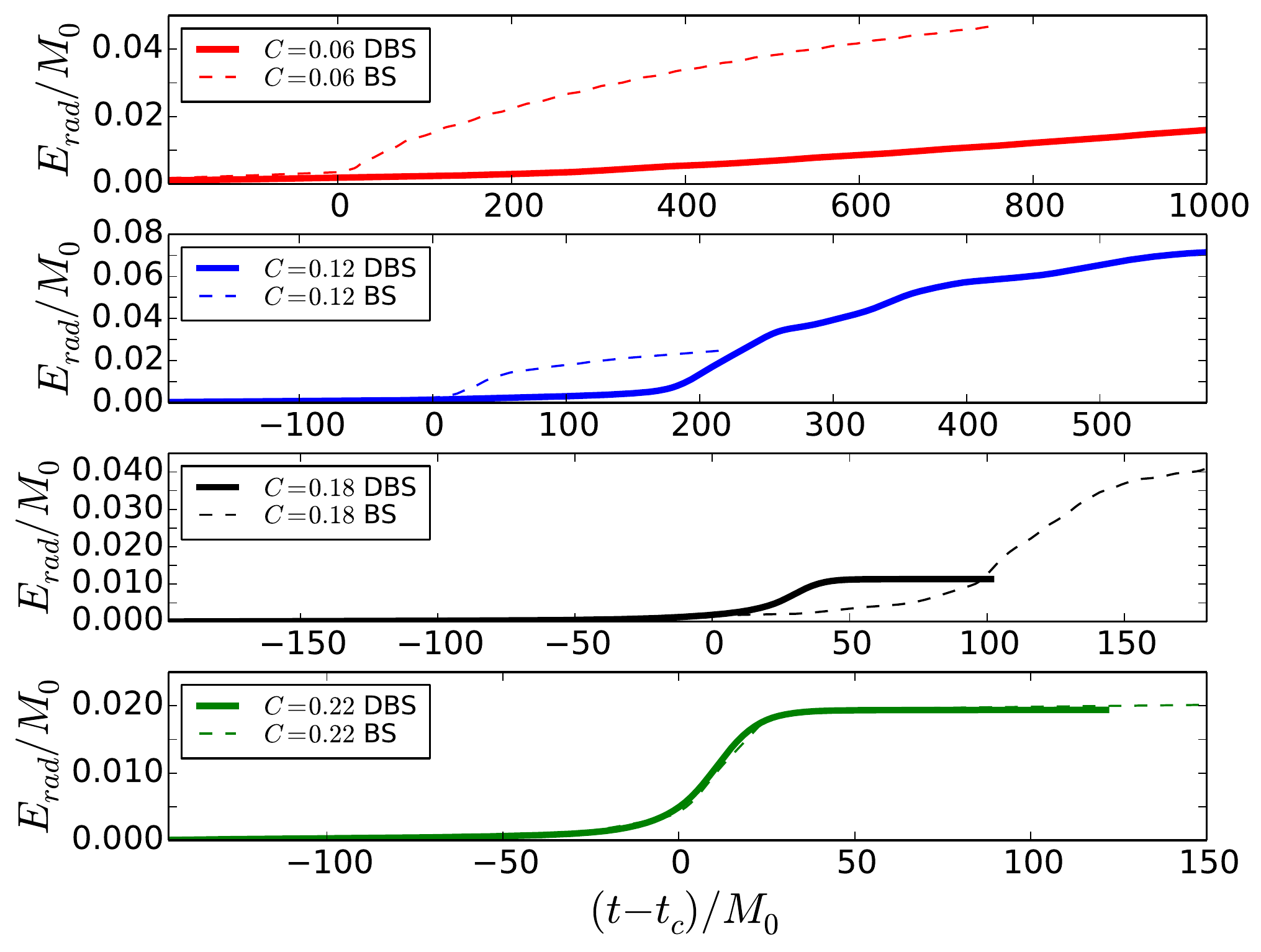}
\caption{ {\em Gravitational waves}. Total GW energy radiated during the coalescence, calculated by integrating in time the quantities displayed in Fig.~\ref{lum2}}
\label{lum5}
\end{figure}

Finally, the luminosity and total radiated energy produced by the main gravitational wave modes $l=2$ are displayed in Figs.~\ref{lum2} and \ref{lum5}, with the specific values  listed in Table~\ref{table2}. The total energy obtained disclose that the cases $C \lesssim 0.12$ are {\em super-emiters}~\cite{Hanna:2016uhs}, since the system emits more than the analogous binary BH system (i.e., about $5\%$ of its initial total mass).  
Following~\cite{palenpani,Hanna:2016uhs}, we can estimate the amount of energy radiated by DBS binaries as
\begin{eqnarray}
{\cal E}^{ac}_{{\rm rad}} \approx
M_0 C,
\end{eqnarray}
which is roughly in agreement with the results of our simulations for the low compact cases (i.e., not collapsing to a BH). The detailed calculation can be found in Appendix~\ref{appA}.

\subsection{DBS versus BSs}\label{comp}

In order to analyze the effect of scalar field interactions, most of the previous plots included not only the analysis of our binary DBS simulations, but also previous binary BS results~\cite{palenpani}. Let us here stress  some of the most significant differences which have not been discussed yet.
  
\begin{figure}
\centering
\includegraphics[width=0.9\linewidth]{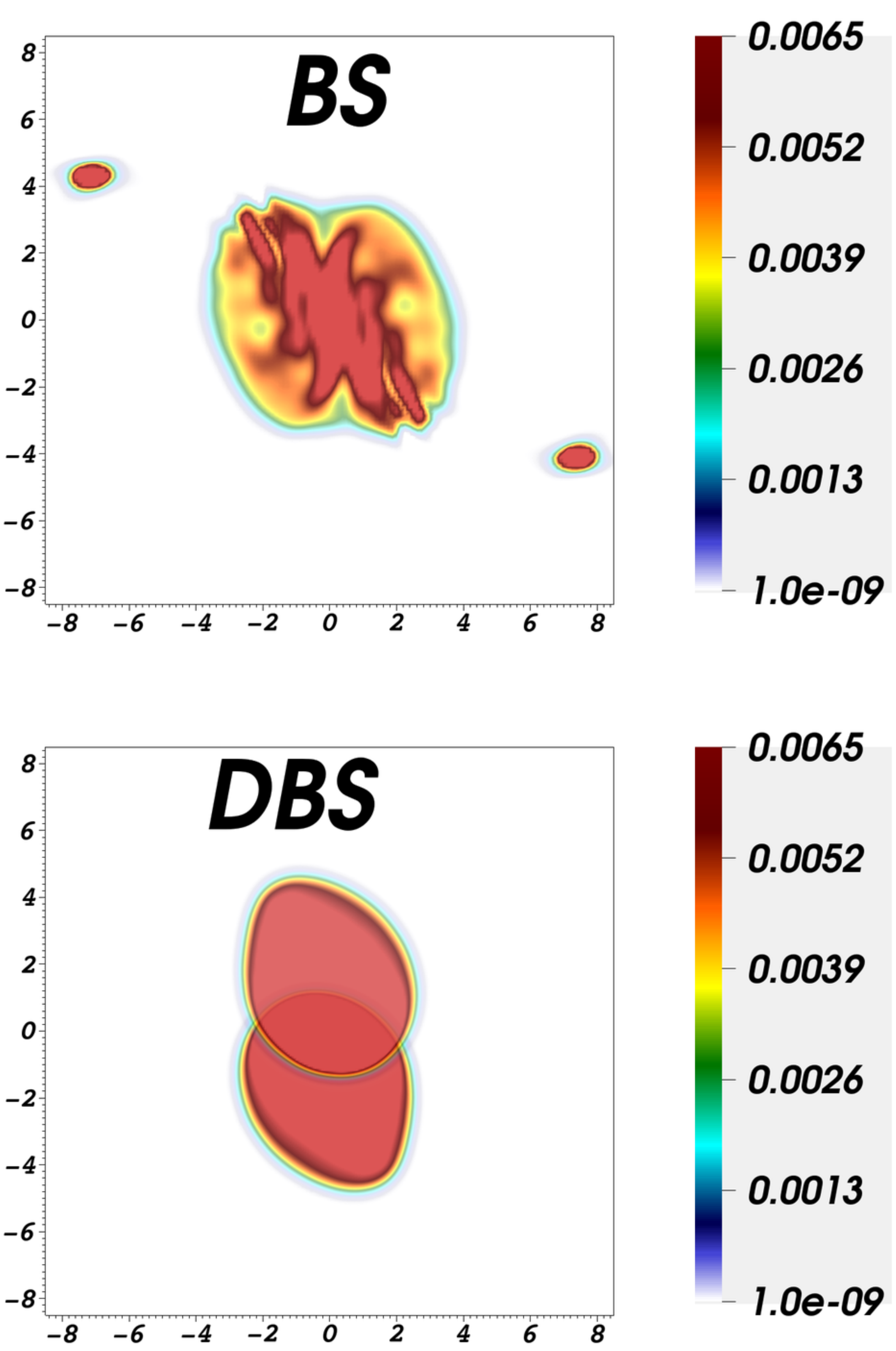}
\caption{ {\em Comparison of DBS vs BSs}. Noether charge densities of DBS and BS binaries, roughly at time $t=t_c + 60$, for the stars with compactness $C=0.12$. Notice that there are two scalar blobs in the BS merger which do not form in DBS collisions.}
\label{comp_blobs}
\end{figure}

An unexpected behavior was recognized in the  binary BSs case with compactness $C=0.12$: the formation of two scalar field blobs which were ejected during the merger, carrying away little mass but an important amount of angular momentum. However, none of the DBS binaries show any evidence of such scalar field blobs. A snapshot at $t=t_c + 60$ of the Noether charge density for DBS and BS binaries with $C=0.12$ is displayed in Fig.~\ref{comp_blobs} to illustrate the different dynamical behavior. The evolution of angular momentum in Fig.~\ref{analysis} also shows considerable losses in the BS case (i.e., resulting into a  sudden decrease after the contact time) as a result of the matter ejection.

It is also quite interesting that, in the BS binary with compactness $C=0.18$ the remnant settles down to a non-rotating BS, while that the remnant of the corresponding DBS binary collapses to a rotating BH. Therefore, one of the effects of matter interactions is to induce an additional pressure that supports the collapse to a BH, increasing effectively the critical transition compactness $C_T$. Consequently, the range of $0.12 < C_{T}\leq 0.18$ valid for DBS is increased to $0.18 < C_{T}\leq 0.22$ for BS.

The gravitational radiation also show interesting differences. During the inspiral phase, the gravitational radiation of DBS and BS binaries are exactly the same, as it is expected. The main difference appears near the contact time. On one hand, binary BS coalescence is governed by scalar field and gravitational forces, which accelerates the dynamics of the system and reduces the time for the remnant to settle down. In this case, after the contact time the two boson stars merge into a rotating BSs which radiate stronger GWs and at a higher frequency than during the inspiral phase. On the other hand, binary DBSs dynamics is driven only by gravitational interactions and there is a smooth  slow transition from inspiral to merger, which can also be appreciated on the GW frequency displayed in Fig~\ref{PPcomp}. 

\section{Conclusions}\label{conclu} 

In this paper we have studied, through full numerical simulations, the dynamics and the gravitational radiation produced during the coalescence of binary DBSs $-$ self-gravitating compact objects composed by bosonic matter that can only interact through  gravity $-$. We have considered four initial compactness, given by $C = \{0.06,0.12,0.18,0.22\}$, of the identical stars forming the binary. For the less compact cases, $C \lesssim 0.12$, the dynamics showed  a smooth transition from inspiral to merger, leading to a superposition of two close-orbiting DBSs. Gravitational radiation emitted on these cases is quite weak during the inspiral phase, although it increases considerably after the contact time. For the highest compactness cases $C\gtrsim 0.18$, the transition from inspiral to merger is quite abrupt. After the merger, the remnant is excessively compact and it inevitably collapses to a BH. GWs on this post-merger stage are given by the classical BH ring-down. 
Therefore, our research reveals that binaries composed by these dark bosonic objects form either; (i) a multi-state BSs for $C < C_{T}$, which emits GW continuously for a long time, or (ii) a rotating BH for $C > C_{T}$ with its typical ring-down signal after the collapse.

We have also compared these DBS mergers with previous simulations of  BS collisions~\cite{palenpani}, where the scalar field behaves as a fluid (i.e., there exists interaction between the stars through both gravity and the scalar field), leading to important differences. Despite presenting the same dynamics during the inspiral phase, scalar field interactions precipitate the plunge of the stars after the contact time and accelerates the dynamics of the remnant. For the DBS binary with compactness $C=0.12$, there is no ejection of blobs, contrary to what happened in the BSs case. Finally, DBS with a compactness $C=0.18$ collapses to a rotating BH after the merger, while the corresponding binary BSs case settles down to a non-rotating BS. 

Finally, the comparison of the instantaneous GW frequencies of DBSs binaries with standard BS binaries and the PN-T4 approximation  reveals, in the cases when the remnant is not a BH, a very particular and distinguishable signature of these compact objects made of dark matter. All these results indicates that there is a new kind of objects, the Dark Stars, with dynamics and gravitational waveforms different and clearly distinguishable from other astrophysical objects like BHs, NSs or even BSs. These distinctive waveforms could be interesting to contrast future observations of aLIGO/aVIRGO.

\vspace{-1pt}
\subsection*{Acknowledgments} 

It is a pleasure to thank Steve Liebling for helpful comments and discussions on the manuscript, as well as Toni Arbona and Carles Bona for useful observations.
CP and MB acknowledge support from the Spanish Ministry of Economy and Competitiveness grants FPA2013-41042-P and AYA2016-80289-P (AEI/FEDER, UE). CP also acknowledges support from the Spanish Ministry of Education and Science through a Ramon y Cajal grant. MB would like to thank CONICYT Becas Chile (Concurso Becas de Doctorado en el Extranjero) for financial support.
We acknowledge networking support from COST Action CA16104 ``GWverse'', supported by COST (European Cooperation in Science and Technology).
%
%
We thankfully acknowledge the computer resources at MareNostrum and the technical support provided by Barcelona Supercomputing Center (AECT-2018-1-0003).

\appendix

\section{ESTIMATE OF THE TOTAL GRAVITATIONAL RADIATION IN THE POST-MERGER STAGE }\label{appA}
Following previous works~\cite{palenpani,Hanna:2016uhs},  we can estimate the amount of energy radiated by DBS binaries during their coalescence. This energy has been calculated for binary BSs in appendix of~ Ref.\cite{palenpani}. Considering the same assumptions for the equal mass case ($M_1=M_2=M$ and $R_1=R_2=R$ ), the energy when both stars make contact can be expressed as:
\begin{eqnarray}
E_\mathrm{contact} = E^\mathrm{pot}_{12} + E^\mathrm{kin}_{12} +
              E^\mathrm{pot}_{1} +  E^\mathrm{pot}_{2} \nonumber \approx - \frac{29}{20} M C.           
\end{eqnarray}
As the merger takes place, the system ultimately settles down into a non-rotating remnant composed by two coexisting DBSs. Therefore, the final energy in the system (beyond the rest mass) is just given by the binding energy, that for a spherical object with uniform density is:
\begin{equation}
E_\mathrm{final} = -\frac{3 M_r^2}{5 R_r}  
         = - \frac{12}{5} M C \frac{R}{R_r}\,,
\end{equation}
where we have considered an upper bound $M_r \approx 2 M$. Assuming no scalar radiation, we can now estimate the radiated energy in gravitational waves radiated after contact ${\cal E}^{ac}_{{\rm rad}}:$
\begin{eqnarray}
{\cal E}^{ac}_{{\rm rad}} = - (E_\mathrm{final} - E_\mathrm{contact}) 
 \approx M_0 C,
\end{eqnarray}
\\
where we have estimated the ratio $R/R_r\approx 1.4$ from our simulations.
Notice that this energy estimate for DBS mergers is almost twice than the one obtained for BSs~\cite{palenpani} (i.e., ${\cal E}^{ac}_{{\rm rad}}\approx 0.48\,M_0\,C$). The main difference comes from the ratio $R/R_{r}$, that in the remnant of BSs mergers was $0.9$ but for DBS mergers is $1.4$.


\bibliographystyle{utphys}
\bibliography{biblio}

\end{document}